# Unidirectional sub-100 ps magnetic vortex core reversal

*Matthias Noske[1], Ajay Gangwar[1,2], Hermann Stoll[1], Matthias Kammerer[1], Markus Sproll[1], Georg Dieterle[1], Markus Weigand[1], Manfred Fähnle[1], Georg Woltersdorf[3], Christian H. Back[2] & Gisela Schütz[1]*

Abstract:

The magnetic vortex structure, an important ground state configuration in micron and sub-micron sized ferromagnetic thin film platelets, is characterized by a curling in-plane magnetization and, in the center, a minuscule region with out-of-plane magnetization, the vortex core, which points either up or down. It has already been demonstrated that the vortex core polarity can be reversed with external AC magnetic fields, frequency-tuned to the (sub-GHz) gyrotropic eigenmode or to (multi-GHz) azimuthal spin wave modes, where reversal times in the sub-ns regime can be realized. This fast vortex core switching may also be of technological interest as the vortex core polarity can be regarded as one data bit. Here we experimentally demonstrate that unidirectional vortex core reversal by excitation with sub-100 ps long orthogonal monopolar magnetic pulse sequences is possible in a wide range of pulse lengths and amplitudes. The application of such short digital pulses is the favourable excitation scheme for technological applications. Measured phase diagrams of this unidirectional, spin wave mediated vortex core reversal are in good qualitative agreement with phase diagrams obtained from micromagnetic simulations. The time dependence of the reversal process, observed by time-resolved scanning transmission X-ray microscopy indicates a switching time of 100 ps and fits well with our simulations. The origin of the asymmetric response to clockwise and counter clockwise excitation which is a prerequisite for reliable unidirectional switching is discussed, based on the gyromode – spin wave coupling.

[1]Max Planck Institute for Intelligent Systems, Heisenbergstraße 3, 70569 Stuttgart, Germany. [2]University of Regensburg, Department of Physics, Universitätsstraße 31, 93053 Regensburg, Germany. [3]University of Halle, Department of Physics, Von-Danckelmann-Platz 3, 06120 Halle, Germany. Correspondence and requests for materials should be addressed to M.N. (email: noske@is.mpg.de).

In thin soft magnetic layers with thicknesses of a few tens of nm and lateral dimensions ranging from 100 nm to 10 µm the vortex structure is the ground state showing an in-plane curling magnetization and a perpendicularly magnetized core at the center with a size of about 10 to 20 nm [1-3]. In spite of its high stability with respect to static external magnetic fields the vortex core can be switched dynamically with low-power sine or pulsed magnetic fields or spin polarized currents [4-7]. This was achieved by resonantly exciting the gyrotropic mode with its eigenfrequency typically in the range from 100 MHz to 1 GHz. By applying rotating in-plane fields this reversal occurs in a unidirectional manner [8-12] since the exciting field couples only to the vortex gyrotropic mode if the sense of rotation of the excitation corresponds to the sense of the vortex gyrotropic mode which is determined by the core polarity. Due to these phenomena the vortex core has been discussed as an extremely stable bit element in digital magnetic random access storage media providing a low-power [7] and selectively addressable switching mechanism with a speed in the range of ns [9]. At much higher (GHz) frequencies vortex structures possess spin wave eigenmodes arising from the magneto-static interaction. Spin wave mediated vortex core reversal was demonstrated experimentally with excitation of multi-GHz rotating field bursts [13]. Hereby unidirectional vortex core reversal was achieved and corresponding micromagnetic simulations indicated a switching time slightly above 200 ps [14, 15] when using one period bursts.

Here we explore, by systematic experimental studies and micromagnetic simulations, how to speed up unidirectional vortex core reversal by excitation with a sequence of two orthogonal monopolar magnetic pulses, less than 100 ps in total duration. The fact that digital pulses can be used for fast spin wave mediated switching instead of rotating field bursts is an attractive aspect for potential technological applications. The phase diagram of vortex core reversal is measured, i. e., the dependence of switching on pulse amplitudes and their durations. A region of unidirectionality is found to be rather large and thus robust against sample dimension variations. The finding that the region of unidirectionality is larger than the corresponding region for rotating field bursts is a further important result. Time-resolved x-ray microscopy movies image the dynamics of the switching process, in excellent agreement with simulations. The measurements indicate a switching time below 100 ps which is in accordance with our simulations. Furthermore, the switching time of unidirectional vortex core reversal by pulsed excitation is simulated in samples of different sizes and the influence of material parameters is studied indicating a lower universal limit of about 70 ps for the switching time. Coupling between spinwaves and the vortex gyromode is identified as the origin of the asymmetric response of the magnetization to clockwise (CW) or counter clockwise (CCW) excitation which is responsible for the unidirectionality in the vortex core reversal by this broad band orthogonal short pulse excitation.

**Results**

We conduct our experiments on Permalloy ($Ni_{80}Fe_{20}$) discs with a thickness of 50 nm and diameters of 490-500 nm. The discs are prepared on top of cross-like copper striplines (cf. top of Fig. 1). By sending current pulses through these crossed striplines [10], magnetic field pulses are generated which excit the Permalloy sample. The excitation consists of two orthogonal monopolar magnetic in-plane field pulses with pulse amplitude $B_0$, pulse length T and a delay of ½ T between the two pulses. The resulting total magnetic field (cf. top of Fig. 1) is **B**(t)=($B_0$ p(t-½T, T), $B_0$ p(t, T), 0) corresponding to CW excitation and **B**(t)=($B_0$ p(t, T), $B_0$ p(t-½ T, T), 0) for CCW excitation, where p(t, T) describes the pulse shape:

$$p(t,T) = \frac{1}{4}(\text{erf}(t*k) + 1) * (1 - \text{erf}((t-T)*k)).$$

erf is the error function and k defines the rise time of the pulses. We define t=0 ps when the first pulse reaches 50 % of its full amplitude as the start of the excitation and t=3/2 T when the second pulse falls below 50 % of its full amplitude as the end of the excitation. Correspondingly the total excitation time is 3/2 T.

To determine the switching phase diagram the core polarity of the vortex structures is measured before and after the pulse excitation using scanning transmission x-ray microscopy at the MAXYMUS endstation at BESSY II, Berlin. For details of the experimental procedure, see Methods section. The experimentally monitored switching behaviour for both vortex core polarities is presented in Fig. 1a-d. The pulse length T is varied from 45 ps to 90 ps and the delay between the two pulses is always tuned to ½ T. This results in a total excitation time ranging from 67 ps to 135 ps. For a CCW sense of rotation (Fig. 1a, b) the pulse $B_x$ in x-direction starts before the pulse $B_y$ in y-direction. For this sense of rotation, it is found that the switching threshold for an initial vortex core 'down' is nearly twice as high as for an initial vortex core 'up'. This asymmetry in switching threshold is nearly independent of the pulse length and proves the undirectionality of the process for pulse amplitudes $B_0$ ranging from 15 mT to 30 mT. Due to symmetry reasons, a vortex core 'up' ('down') excited with a CW sequence is expected to have the same switching behaviour as a vortex core 'down' ('up') excited by a CCW sequence. This is checked and confirmed for the shortest pulse lengths by inverting the timing of the pulse sequence so that $B_y$ started before $B_x$ corresponding to a clockwise excitation. Within the accuracy of the measurements the switching threshold found for core `up` at CW excitation (Fig. 1c) is the same as for core `down` at CCW excitation (Fig. 1b) and the threshold for core `down` at CW excitation (Fig. 1d) is the same as for core `up` at CCW excitation (Fig. 1a). Due to this agreement in combination with the symmetry argument this check is omitted for longer pulse lengths. The measurements shown in Fig. 1 further demonstrate that highly reliable and reproducible unidirectional switching is possible since in the connected areas in the phase diagram representing switching (non-switching) events not a single non-switching (switching) event is found. Additionally, the pulse lengths of only 45 ps and total excitation times of 67 ps are the shortest values shown for unidirectional vortex core reversal. Three-dimensional micromagnetic simulations are performed with the OOMMF code [16] (for simulation parameters see Methods section). The results are shown in Fig. 1e, f and demonstrate a good qualitative agreement with the experimentally observed asymmetry in switching thresholds. For high pulse amplitudes a region exists in the phase diagram (Fig. 1a, shaded area) where the vortex core polarity after excitation is the same as before. Analyzing the corresponding simulations of Fig. 1e, this is explained by double switching events. For further increased amplitudes even triple switching is observed.

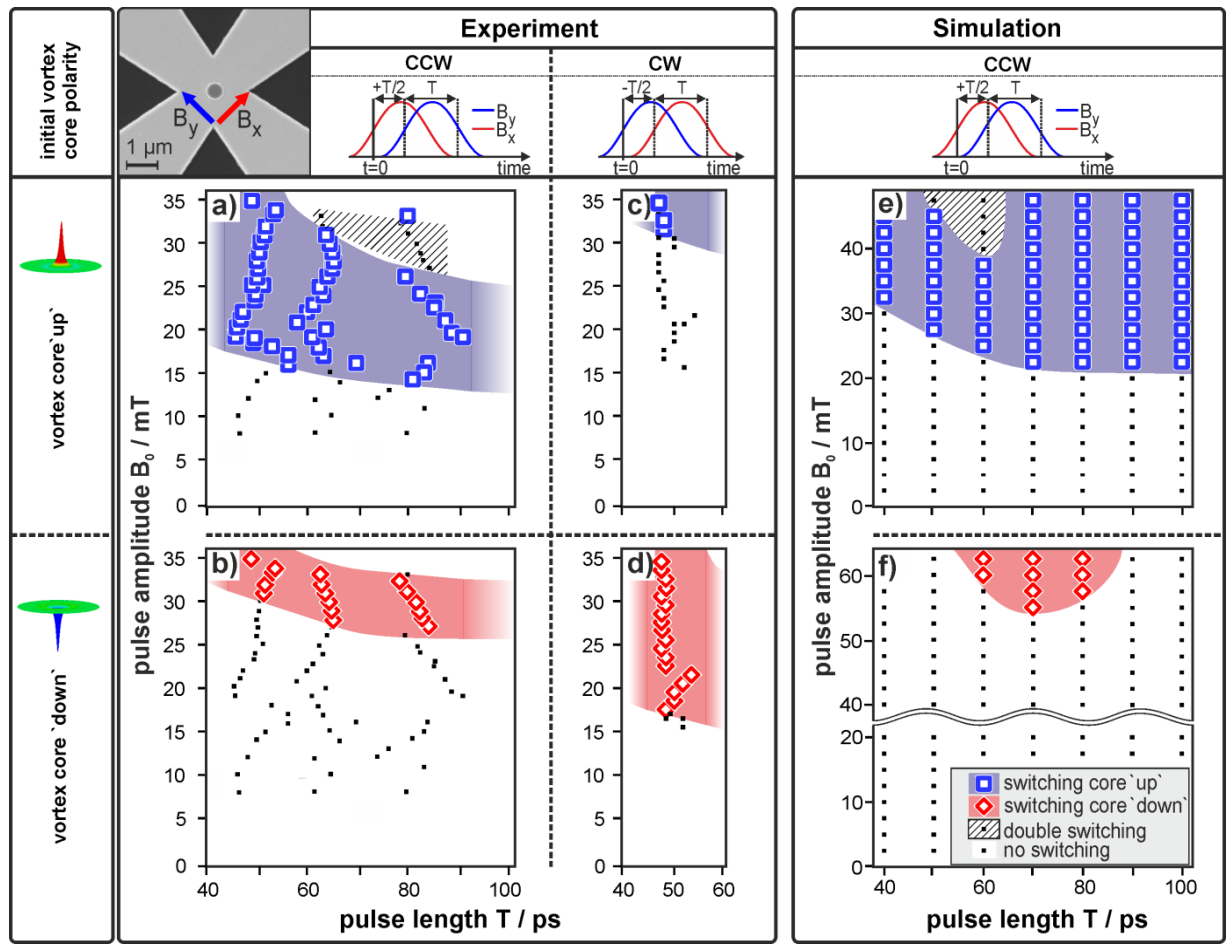

Fig. 1: Phase diagrams for vortex core reversal by pulsed excitation obtained experimentally (sample A: a – d) and by micromagnetic simulations (e, f). A scanning electron microscope image of the crossed striplines with the Permalloy disc is shown at the top. The arrows indicate the directions of the magnetic field pulses that produce the CCW/CW excitation sequence. Unidirectional reversal is found in a broad region, where CCW excitation switches a vortex with core up (a, e) but not a vortex with core down (b, f). For CW excitation, this behaviour is inverted (c, d) as is expected for symmetry reasons.

It has to be pointed out that the switching thresholds in our experiments are found to be systematically lower by about 33 % compared to the simulations. This tendency, which has been observed before in the case of core reversal by static out of plane fields [17], seems to be a systematic phenomenon. It may be partly be explained by sample defects acting as nucleation centers for the reversal process and by thermal excitations, which are not accounted for in the simulations. Additionally, in micromagnetic simulations the treatment of a Bloch point which is present during the reversal process is problematic and can increase the switching threshold as discussed by Thiaville et al. in detail in [17]. It might be of considerable technological interest that the experimental switching thresholds are systemically lower than expected from the simulations as the required fields can be realized easier.

**Spin wave mediated vortex core reversal.** To get more insight into the dynamics of the reversal process the out-of-plane component of the magnetization during the switching process is imaged by stroboscopic time-resolved scanning transmission x-ray microscopy and is compared with the result from a corresponding micromagnetic simulation (Fig. 2).

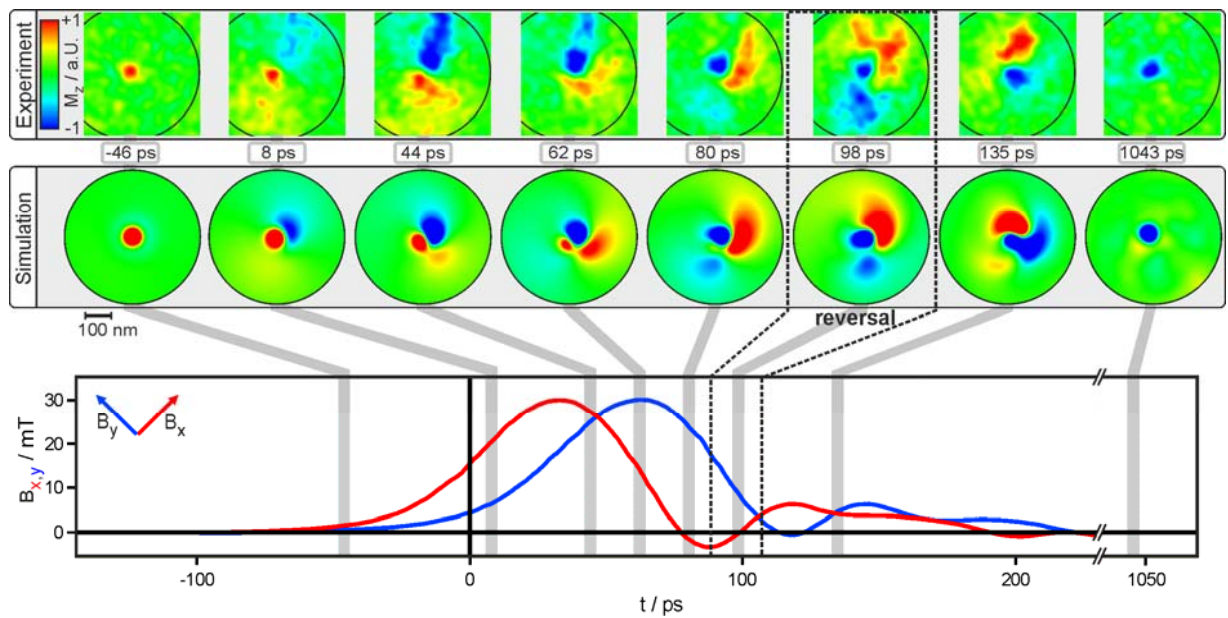

Fig. 2: Time-resolved imaging of the VC reversal process by the excitation with orthogonal field pulses of 30 mT strength, pulse width of 60 ps and a delay of 30 ps (shown in the lower row). The top row shows the dynamic response of the out of plane magnetization measured with time-resolved STXM (sample B). The middle row shows the z-component obtained from the corresponding micromagnetic simulation averaged over the thickness and convolved with the experimental temporal and lateral resolution.

The exciting pulses have a length of T=60 ps, a pulse amplitude of B=30 mT and a delay of 30 ps between the two pulses corresponding to a CCW excitation. The experimental excitation is shown in Fig. 2 and is also used for the simulation (for simulation parameters see Methods section). Note that this experimental pulse shape deviates slightly from the ideal shape given by **B**(t)=(*B* p(*t, T*), *B* p(*t-½ T, T*), 0). However, no significant deviations are found when repeating the simulation with the ideal shape. The simulation is convolved with the experimental lateral and temporal resolution. Measurements and simulations show excellent agreement. The vortex core initially points up (red dot). During the rising edge of the first pulse, a bipolar structure forms similar to an n=1 m=+1 spin wave [13]. The counter clockwise rotation of this structure follows the rotational sense of the exciting magnetic field. From t≈45 ps to t≈85 ps the downwards oriented magnetization (blue) gets localized close to the position of the original vortex core forming a 'dip' while the core smears out and its contrast reduces. The latter can be explained by a compression of the vortex core in combination with tilting of the core from the z-direction. The measurement indicates core reversal at about t=100 ps. From the (unconvolved) simulations it can be deduced that at t=85 ps the switching process begins and is completed at t=106 ps when the original vortex core 'up' has switched to a vortex core pointing 'down' (blue spot) in all discretisation layers of the three-dimensional simulation. To characterize the speed of the reversal we define a switching time $\theta$ which measures the time from the start of the excitation until the core is completely reversed. In this case the switching time $\theta$ is 106 ps.

**Speed of the reversal process.** When thinking about a possible application of the vortex core as fast storage element it might be of interest how this switching time $\theta$ depends on sample parameters and the excitation. Additionally, to achieve higher storage densities smaller elements are favorable. Therefore the switching times as a function of pulse amplitude and pulse length (T=40 ps, T=60 ps, T=90 ps) are deduced from the micromagnetic simulations shown in Fig. 1 (for a diameter $d_1$=500 nm) and for two additional Permalloy platelets with smaller diameter ($d_2$=250 nm, $d_3$=125 nm), all with a thickness of 50 nm. The results are shown in Fig. 3. The switching time strongly depends on the pulse amplitude but surprisingly is nearly independent of the pulse length. Furthermore, smaller disks below diameters of 500 nm do not allow shorter switching times in contrast to the prediction of [14].

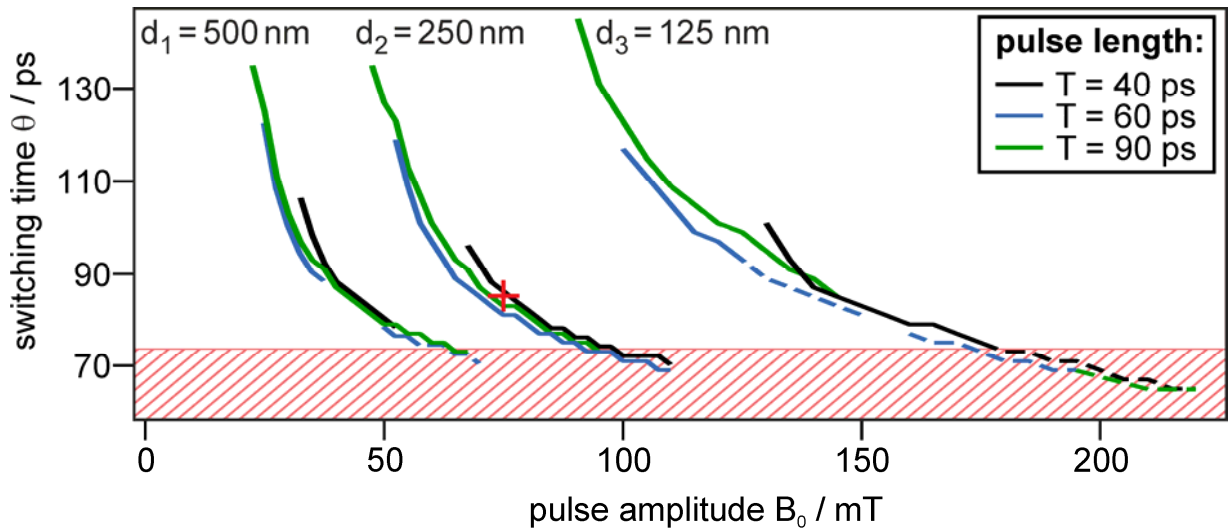

Fig. 3: Switching time as a function of pulse amplitude B derived from micromagnetic simulations. Results for three samples of different diameter (d=500 nm, d=250 nm, d=125 nm) and three different pulse lengths T are shown. Dashed lines indicate where the unidirectionality is lost since switching occurs for both vortex core polarities. The switching time only slightly depends on the pulse length. For all three samples unidirectional switching within 70-75 ps is possible.

When scaling the pulse amplitudes with the inverse diameter (500 nm disc: B'= $d_1/d_1$ $B_0$; 250 nm disc: B'=$d_2/d_1$ $B_0$; 125 nm disc: B'=$d_3/d_1$ $B_0$), the switching time shows similar dependence of this scaled pulse amplitude B' for all three samples. Unidirectional core reversal is possible with switching times as short as 75 ps in all cases. Higher pulse amplitudes would allow even shorter switching times, however, the unidirectionality is lost since switching occurs for both core polarities (indicated by the dashed lines). Especially in the smallest sample, the pulse length has to be chosen appropriately since there the region of unidirectionality depends most strongly on the pulse length. To estimate the influence of the remaining parameters (sample thickness, material properties, cell size of the simulation), a single parameter is varied separately starting from the simulation marked by the red cross in Fig. 3 ($d_2$=250 nm, B=75 mT, T=40 ps, switching time $\theta$=86 ps). It is found that the saturation magnetization has the strongest influence on the switching time $\theta$ which can be easily explained. A reduction from $M_S$=830 kA/m to $M_S$=660 kA/m increases $\theta$ from 86 ps to 100 ps due to the fact that less energy is coupled into the system by the exciting pulses leading to a later start of the reversal process. A reduction of the thickness from 50 nm to 40.625 nm reduces the switching time to 80 ps while increasing the thickness to 59.375 nm increases the switching time to 100 ps. Here for the three different thicknesses the reversal process starts at t=70 ps, but the actual reversal takes longer for thicker samples. Varying the damping constant in the range from $\alpha$=0.001 to $\alpha$=0.01 or the exchange constant in the range from $A_{Ex}$=11 * $10^{-12}$ J/m to $A_{Ex}$=15 * $10^{-12}$ J/m has no detectable influence on the switching time. Reducing the cell size of the simulation from (3.215 nm)³ to (2 nm)³ increases the switching time by 4 ps which is probably related to the Blochpoint present during the reversal process [17, 18].

**Discussion**

In the final part we discuss the physical origin of the unidirectionality of vortex core reversal with very short orthogonal magnetic pulses. It differs fundamentally from the excitation with a rotating magnetic field with a well-defined frequency [13]. There the frequency can be tuned to a specific CW or CCW spin wave mode eigenfrequency and excitation only occurs, if both the frequencies and the senses of rotation of the external field and this specific spin wave mode are the same. Only if this is the case, energy is resonantly coupled into the system and vortex core reversal occurs. In contrast, the very short pulses used in the present paper show a several GHz broad frequency spectrum and thus frequency tuning to a specific CW or CCW vortex core eigenmode fails. Nevertheless, a unidirectional vortex core reversal is found for such a short and broadband excitation (cf. Fig. 1), in spite of the fact that our micromagnetic simulations reveal that the energy coupled into the system is nearly the same for CW and CCW excitation.

We suggest assigning the physical origin for this CW/CCW asymmetry for short pulse excitation to a coupling of the spin waves with the gyrotropic mode. Such coupling is present as otherwise the well-known frequency splitting of CW and CCW rotating spin wave modes would not exist [21, 22]. For vortex polarity 'up' (p=+1) the vortex gyromode only shows a CCW sense of rotation whereas for vortex core polarity 'down' (p=-1) only a CW sense of rotation is observed. This is also well-known and can be calculated analytically using the Thiele equation [19, 20]. As the spin wave modes are coupled to the gyromode the response of the magnetization to CW or CCW excitation is different depending on whether the sense of rotation of the excited spin wave mode is the same or opposite to the sense of rotation of the gyromode.

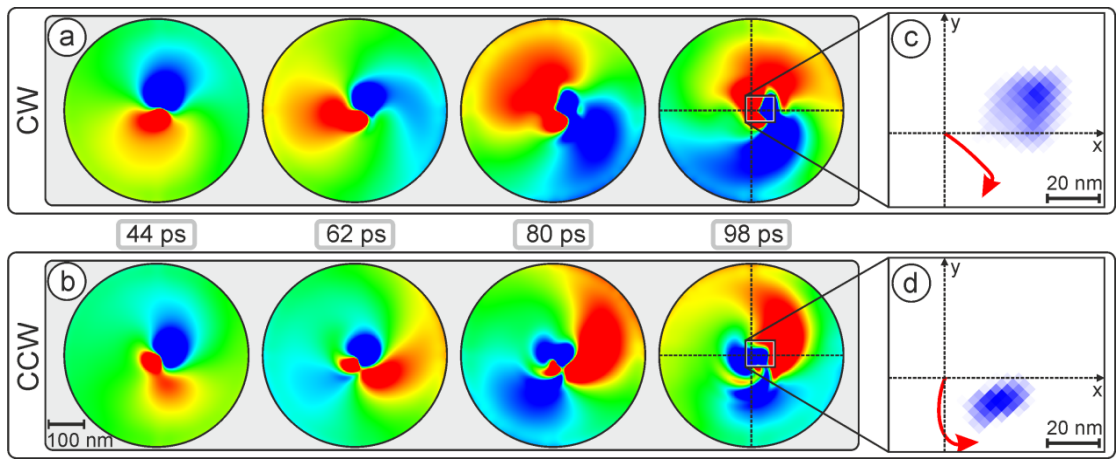

Fig. 4: Comparison of the simulated evolution of the out-of-plane magnetization for the two rotation senses of the excitation. The different evolutions of the magnetization profiles for the CW- and CCW-case are displayed in the consecutive snapshots (a, b) for an initial vortex core 'up' (the CCW-case corresponds to Fig. 2a, but is not convolved with the experimental resolution here). The core-trajectories are shown on the right side (c, d) by the red arrows. Their origin corresponds to the equilibrium position of the core, x=y=0. In the blue regions marked in c, d the downwards oriented dip forms. These regions are obtained by averaging all cells with $M_z < -0.9\ M_s$ over the sample thickness and from t=0ps to t=100ps. In the CCW case this dip region is closer to the original vortex core and more localized than in the CW case. Also, in the CCW case the core moves towards this dip region, while it moves away from it in the CW case leading to vortex core reversal for CCW excitation only.

Micromagnetic simulations in Fig. 4 reveal the differences in vortex dynamics for CW and CCW excitation (pulse length T = 60 ps) which lead to the observed asymmetric switching thresholds. The vortex has a core pointing 'up' (polarity p=+1) and a clockwise in-plane magnetization (circulation c=-1). Other combinations of c, p and sense of the exciting fields are not discussed since they can be derived from the two cases shown here by simple symmetry considerations (see notes [23]). For a CW pulse sequence, mainly the n=1/m=-1 mode (eigenfrequency f=7.1 GHz) is excited while for CCW excitation mainly the n=1/m=+1 mode (eigenfrequency f=9.5GHz) is excited. Due to the coupling with the vortex gyromode, excitation of the spin waves also leads to a movement of the vortex core. Both the simulated magnetization profile (Fig.4a, b) and the trajectory of the vortex core (Fig. 4c, d) are clearly different leading to core reversal for CCW excitation only. The difference in switching threshold can be explained by these non-symmetric vortex core trajectories. Guslienko et al. [24] explained the formation of the out-of-plane magnetized dip region preceding the vortex core inversion by the introduction of an effective gyrofield caused by the moving vortex core. In our case, the different core trajectories (Fig. 4c, d) lead to different gyrofields and therefore different dip formations. For both CW and CCW excitation the formation of the dip is located left to the vortex core movement (Fig. 4c). But in the CCW case the dip forms closer to the core (Fig. 4d), is more localized and partly enclosed by the core's trajectory. This results in core reversal where the dip is transformed into a vortex core pointing down while the original core is annihilated. In the CW case, the average distance between the created dip and the original vortex is larger and the dip is not confined by the core's trajectory. The dip partly splits up into a vortex-antivortex (VA) pair followed by the annihilation of the original vortex core. However, this process does not occur across the whole thickness of the disc. In contrast, the splitting into VA pair starts at the bottom layer of the three-dimensional simulation, and progresses only to approximately one third of the disc thickness. Subsequently, the remaining dip in the upper simulation layers dissolves, starting at the top layer. Thus the reversal process is not completed and the original vortex core 'up' remains. This is in contrast to previous investigations [4, 5, 7, 8, 13, 14, 18] where the formation of a VA pair led to core reversal.

In conclusion, we have experimentally demonstrated high-speed unidirectional vortex core reversal by orthogonal monopolar short pulses with pulse lengths between 90 ps and 45 ps. We observe a unidirectional switching behavior in a broad range of field amplitudes and pulse lengths, in qualitative agreement with simulation. The asymmetry in the response to clockwise and counter clockwise excitation can be explained by the coupling between the gyrotropic mode and the spin waves resulting in a different evolution of the magnetization for opposite rotation senses. We imaged the spin wave and the vortex core evolution by time-resolved x-ray microscopy in comparison with micromagnetic simulations, the results of which agree excellently with the experimental findings. As indicated in the x-ray movies and confirmed by the simulations a switching time of about 100 ps can be achieved. The possibility to further reduce the switching times for smaller sample diameters is explored by the corresponding micromagnetic simulations, where a lower nearly size-independent limit of about 70 ps is deduced. These experimental and simulation studies provide important information in view of the use of vortex core as fast, stable and low-power storage bit: The vortex core switching times can be reduced below 100 ps and are thus in the order of the fastest electronically induced magnetization reversals in nanostructures [25] or even below. The unidirectionality covers a wide range of excitation amplitude and pulse duration and is therefore robust for arrays with irregular sample geometries. The ultimate value of switching times of about 70 ps cannot be reduced by decreasing the vortex size, whereby the optimum combination between size – determining the package density - and required field amplitude can be chosen for a given application.

**Methods**

**Experiment.** Permalloy ($Ni_{80}Fe_{20}$, 50nm thick) discs with diameters of 490 nm (sample A) and 500 nm (sample B), are thermally evaporated on top of cross-like copper (Cu, 150 nm thick) striplines with a width of 1.6 µm (cf. top of Fig. 1). To avoid surface oxidation, the discs are capped with a 4 nm Al layer. The samples are structured in multiple steps on top of silicon nitride membranes using a combination of optical and electron-beam lithography and lift-off processes. By sending current pulses through the crossed striplines, magnetic field pulses up to a field strength of 35 mT can be generated to excite the Permalloy sample. A sampling oscilloscope with a bandwidth of 50 GHz is used to monitor the exciting signals allowing the calculation of the magnetic field strength at the location of the sample [10]. The vortex structures are investigated by scanning transmission x-ray microscopy at the MAXYMUS endstation at BESSY II, Berlin, which provides a lateral resolution of 25 nm and high magnetic contrast based on the x-ray magnetic circular dichroism effect [26] at the Ni $L_3$-edge. A sophisticated lock-in-like data acquisition scheme allows low-noise time-resolved pump-and-probe measurements of the magnetization dynamics of the vortex structure with a temporal resolution down to 45 ps. The complete pump-and-probe cycle for the measurement of the reversal process shown in Fig. 2 consists of the CCW pulse sequence shown in the figure and an additional CW pulse sequence at t=6.5 ns that resets the vortex core its initial 'up' polarity (not shown in Fig. 2). The stroboscopic experiment is repeated with a frequency of 75 MHz over more than $10^{11}$ cycles to obtain the required signal to noise ratio.

**Micromagnetic simulations.** Three-dimensional micromagnetic simulations are performed using the object-oriented micromagnetic framework (OOMMF) [16]. Unless otherwise stated, disc shaped platelets with a diameter of 500 nm and a thickness of 50 nm with cubic simulation cells of (3.125 nm)³ are used. The damping constant $\alpha$=0.0077 and saturation magnetization $M_S$=830 kA/m are derived from the damped vortex core gyration after exciting sample B with a magnetic monopolar pulse and measured using time-resolved STXM. Typical values for Permalloy for the exchange constant ($A_{Ex}=13 * 10^{-12}$ J/m) and the gyromagnetic ratio ($\gamma=2.21 * 10^5$ m/As) is used. The anisotropy constant is set to zero ($K_1=0$). The position of the vortex core in Fig. 4c, d is tracked by fitting a Gaussian to the z-component of the magnetization.